# Disruptive Atomic Jumps Induce Grain Boundary Stagnation

Xinyuan Song, Chuang Deng*

Department of Mechanical Engineering, University of Manitoba, Winnipeg, MB R3T 2N2, Canada

* Corresponding author: Chuang.Deng@umanitoba.ca

## Abstract

Grain growth in polycrystalline materials can be impeded by grain boundary (GB) stagnation. Using atomistic simulations, we observed that during GB migration, the disruptive jumps of a few GB atoms can disturb the original ordered collective movement of GB atoms, leading to the stagnation of the entire GB. These disruptive atomic jumps can be activated by both high driving forces and high temperatures, with even jumps of a few atoms capable of causing the stagnation of an entire GB. This mechanism also explains the non-Arrhenius behavior observed in some GBs. Additionally, a large model size could increase the rate of disruptive atomic jumps, and a clear transition in thermal behavior is observed with the increase of the GB size in GBs exhibiting clear thermally activated stagnation. Our further investigation shows that the disruptive atoms involved in these jumps do not differ from other GB atoms in terms of atomic energy, volume, density, local entropy, or Voronoi tessellation, and no "jam transition" was observed in the energy barrier spectra. This fact makes those disruptive jumps challenging to detect. To address this issue, we propose a displacement vector analysis method that effectively identifies these subtle disruptive jumps.

## 1. Introduction

Most solid materials, especially for metals, are polycrystalline, consisting of grains with varying orientations. The grain boundaries (GBs) are interfaces between these grains. During heat treatment or mechanical deformation, GBs migrate to minimize the excess energy introduced by the GBs or to alleviate stresses. Theoretically, given sufficient relaxation time, the grains in a pure material can continuously grow until the material becomes a perfect single crystal, representing the configuration of minimum energy. However, in practice, GBs often encounter impediments after a period of migration, which leads to a cessation of grain growth. This phenomenon is known as GB stagnation.

Previous studies on GB stagnation are mainly centered around solute or impurity drag[1–7]. However, this explanation is not always sufficient, as GB stagnation can also occur in pure materials [8,9]. Fan et al.'s investigation [5] demonstrates that the distribution of grain size is unaffected by solute drag, remaining the same as observed in pure materials. Frost et al.'s study [4] shows that the solute drag effect is only evident when the driving force of the GBs is small. Li et al.'s study [10] also shows that the solute drag alone is insufficient to halt the GB migration. Gottstein et al.[11–14] modified Von Neumann-Mullins Relation [15,16] by considering the triple-junction drag, and Barmak et al. [7]'s two dimensional phase field simulations shows that with the increasing triple-junction drag, the distribution of the grain size is more closely aligned with the experimental data. Other contributing mechanisms include the Zener pinning effect, influenced by structures such as GB grooving [17–19] and second phase particles [20,21].

Holm and Foiles [22] proposed that GBs themselves can induce GB stagnation in the absence of impurities. They analyzed 388 GBs from the Olmsted database and discovered that, prior to the roughening transition, smooth GBs typically exhibit low mobility. Remarkably, a small fraction of these smooth GBs can lead to the stagnation of the entire GB network. This finding contradicts traditional pinning theories, which assert that a substantial proportion of pinning sites on the GB surface is necessary to induce stagnation [21,23,24]. Other potential self-caused mechanisms might involve GB phase transitions, a phenomenon widely observed both in simulations [25–27] and experiments [28–30]. Homer et al.'s study [31] demonstrates that the mobilities of the meta-structures within the same GB can vary significantly. Thus, if a GB transitions into a meta-structure with reduced mobility during its migration, it could potentially cause GB stagnation.

In this study, we reveal a mechanism whereby disruptive atomic jumps within the GB can impede GB migration, ultimately leading to GB stagnation without causing detectable GB structural phase transitions. These disruptive atomic jumps may be triggered by high driving forces or elevated temperatures. This mechanism provides a possible explanation for GB stagnation in impurity-free materials and accounts for the non-Arrhenius migration behavior observed in some GBs.

## 2. Methods

### *2.1 Modeling*

In this study, we investigate the Σ15 (2 1 1) Ni GB (p14 in the Olmsted database [32]). The model's dimensions and orientation parameters are detailed in Fig. 1. The boundary conditions are periodic in the *y* and *z* directions (within the GB plane) and have free surfaces in the *x* direction. Initially, the model is expanded at different temperatures using the thermal expansion coefficients. It then reaches equilibrium under the isothermal-isobaric (NPT) ensemble for 10ps, followed by a 5ps annealing in the canonical (NVT) ensemble using Berendsen thermostat [33]. Simulations were conducted using the Large-scale Atomic/Molecular Massively Parallel Simulator (LAMMPS) [34] with an embedded atom method (EAM) potential tailored for Ni [35]. The simulation results are visualized through the Open Visualization Tool (OVITO) [36].

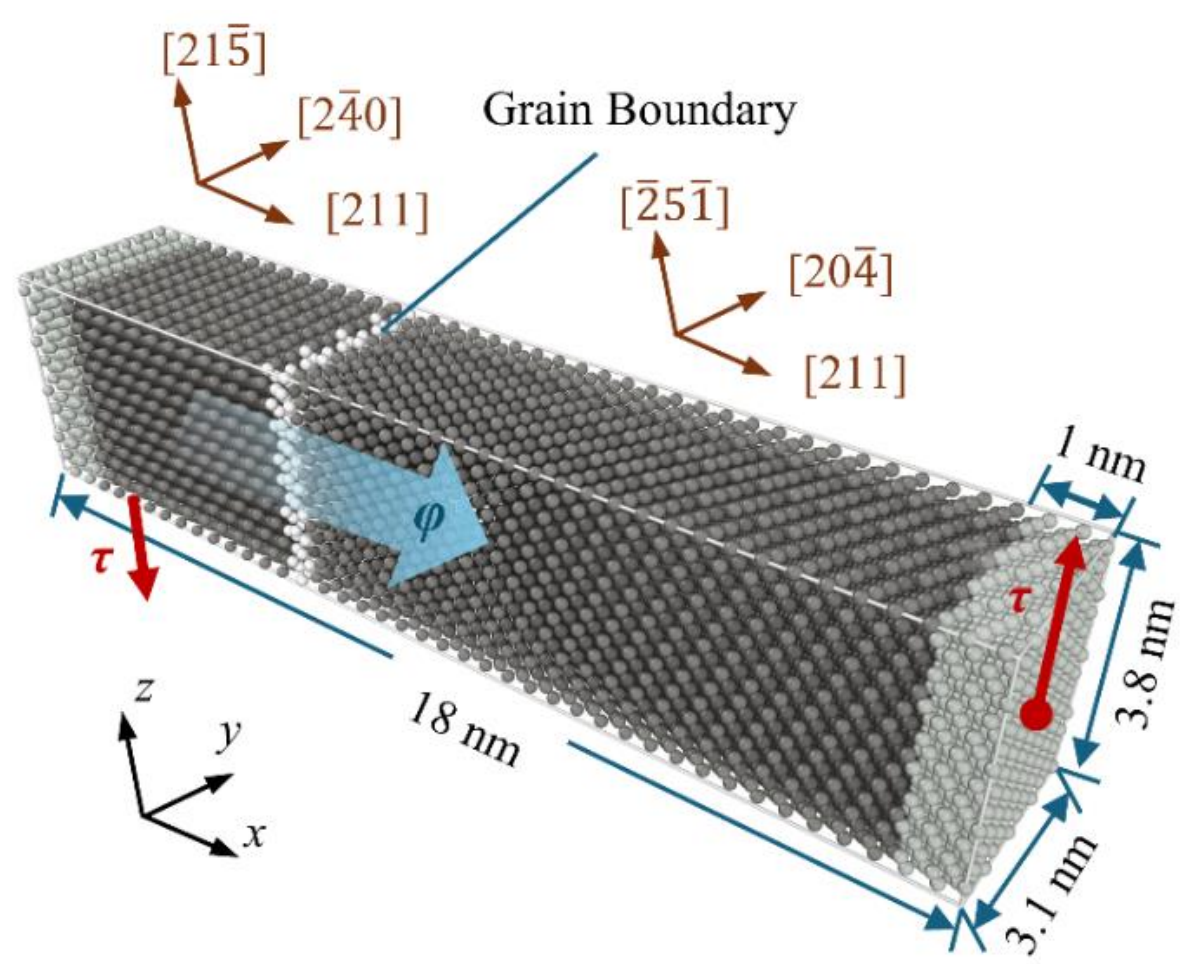

Figure 1 The dimensions and orientation parameters of the Σ15 (2 1 1) Ni GB model, and different types of driving forces applied.

*2.2 GB migration simulations*

In the force driven GB migration simulations, two types of driving forces are applied: shear stress ($\tau$) and normal driving force ($\varphi$). Shear stress is applied by exerting opposing forces in the z-direction on the slabs located at both ends of the model (the light gray areas in Fig. 1). The normal driving force is introduced through an energy-conserving orientational (ECO) force [37,38], manifesting as a synthetic energy jump across the GB, as shown in Fig. 1. The position of the GB is tracked through the order parameter [37].

In determining the mobility and thermal behaviour of GBs, we carried out random walk simulations [39–41], in which the position of GB fluctuate randomly under zero driving force and the microcanonical (NVE) ensemble. Here, we utilized the Fast Adapted Interface Random Walk (FAIRWalk) method [39] to calculate GB mobilities. This approach can significantly reduce computational costs, especially effective for large models later investigated in our study. We conducted simulations of 5 ns each, repeating them 20 times at each temperature with different initial velocity distributions. To enhance statistical evaluation of the variance, we applied the ergodic principle by dividing each 5 ns dataset into ten 500 ps segments. This approach yielded 200 independent data points at each temperature. The calculation method for GB mobility and statistical reliability testing are detailed in the literature [39].

*2.3 Energy landscape mapping*

Two energy landscape mapping techniques are employed: the nudged elastic band (NEB) method [42–44] and the activation-relaxation technique (ART nouveau) [45,46]. The NEB method is known for its effectiveness and precision in identifying the minimum energy path between two configurations. In our simulations, we applied a nudging force of 1.0 eV/Å in both parallel and perpendicular directions, as per Mares et al.'s method [44]. Subsequent comparisons between molecular dynamics (MD) and NEB results demonstrate the effectiveness of this approach in our study of disruptive jump within GB area.

The ART nouveau method, on the other hand, can explore saddle points on the energy surface surrounding the current state without requiring predefined final configurations. This technique involves randomly moving a few atoms in the system at a time; if the resulting configuration is stable (a local minimum on the energy surface), it automatically calculates and outputs the energy barrier for that movement. This provides a more localized approach to energy landscape exploration, focusing on changing the local area rather than the entire GB structure, and has been utilized in previous studies of GB migration [47,48]. In our study, we set an eigenvalue threshold of -1 for the Hessian matrix in the Lanczos algorithm [49] to determine whether the configuration exits the current energy basin, and a force threshold of 0.05 eV/Å to achieve convergence at the saddle point.

## 3. Results

### *3.1 Driving force induced GB stagnation*

Grain growth can be driven both mechanically and thermally. Initially, we investigated GB stagnation induced by driving forces, without any detectable structural transitions. Various types and magnitudes of driving forces were applied to the GB at a low temperature of 100K. Figure 2 demonstrates that with the increase of the driving force, the velocity of the GB (reflected by the slope of the curves) gradually increases. However, once the driving force surpasses a specific threshold, e.g. 88.7 MPa, the GB stagnates after migrating a certain distance. This behavior is consistent across both shear stress and normal driving force. A similar phenomenon has previously been reported in the literature [50]. It is worth noting that the GB stagnation observed in this study differs from the previously reported "stick-slip" GB migration pattern [51], where temporary GB stagnation and movement alternate. In our case, the GB stagnation is of a long-term nature, potentially leading to permanent changes in the GB structure. Additional long-term simulation results at varying magnitudes of driving force are presented in Supplemental Fig. S1.

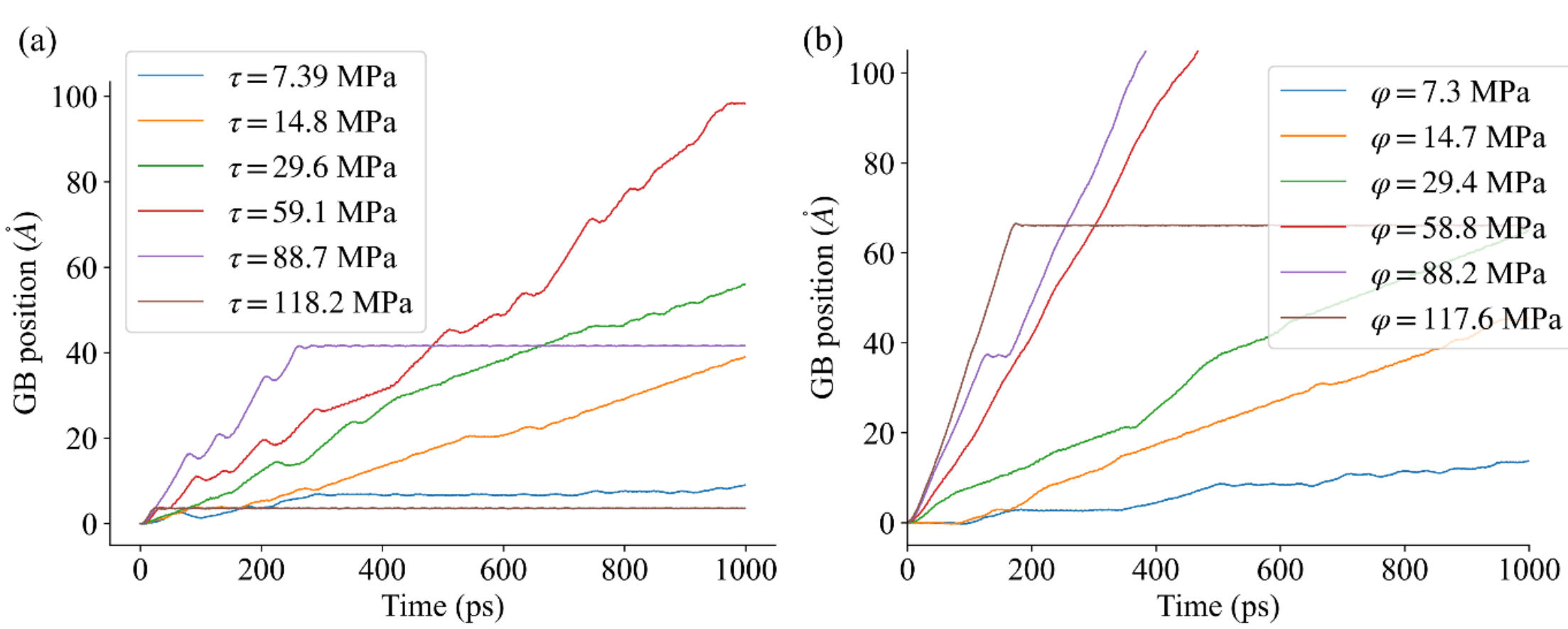

Figure 2 Plots of GB position vs. time curves under different (a) shear stresses $\tau$ and (b) normal driving forces $\varphi$ at 100K.

To determine the causes of GB stagnation, we analyzed a case under a shear stress of 118.2 MPa at 100K. Figure 3 illustrates that prior to stagnation (during normal GB migration), the GB atoms are arranged in an ordered pattern and move collectively (as shown in Fig. 4a), maintaining the GB structure unchanged during migration. After the GB stagnation, a delayed area appears on the GB surface where the atomic arrangement becomes relatively chaotic. Apparently, this area exhibits lower mobility compared to other regions of the GB.

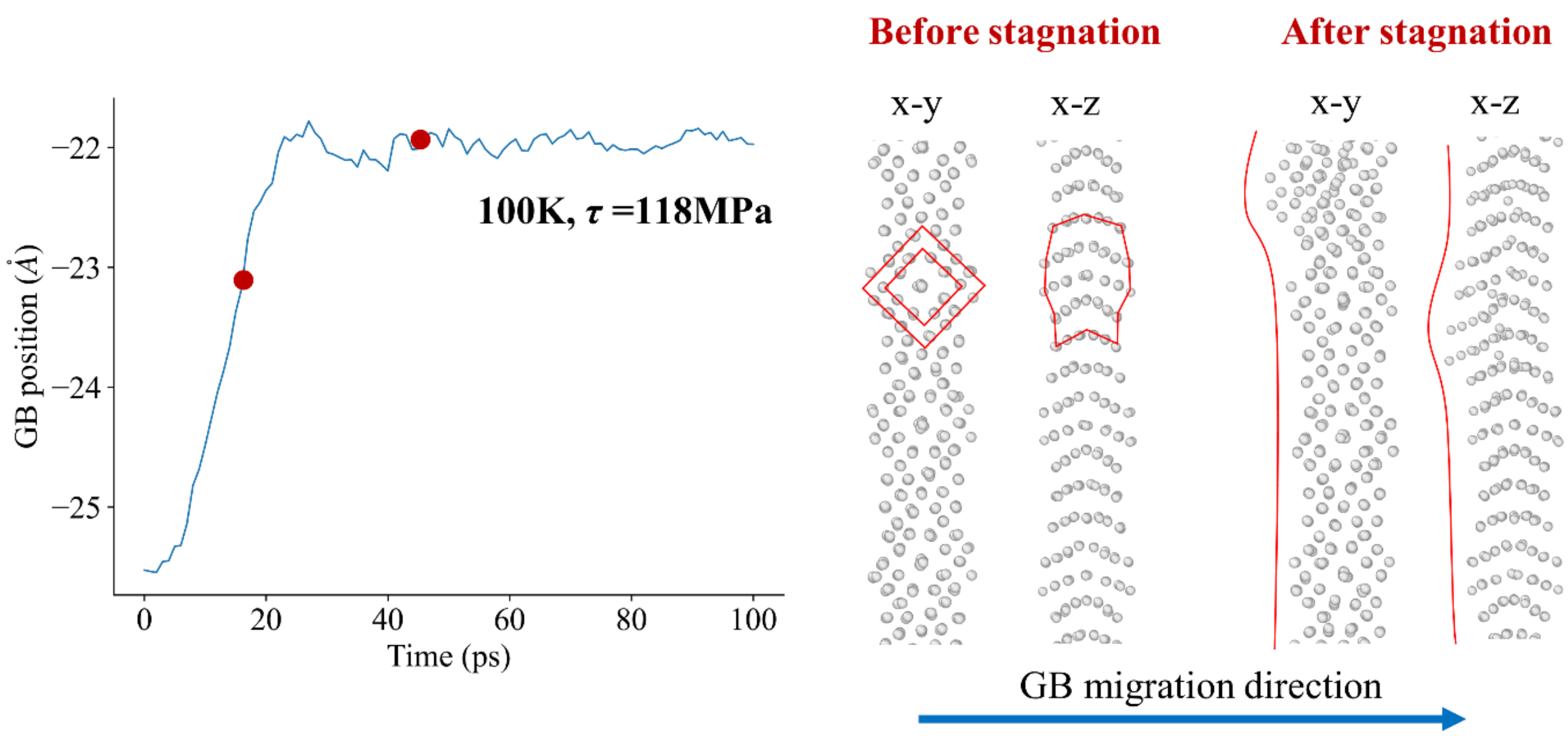

Figure 3 The arrangement of GB atoms before and after the GB stagnation

Further frame-by-frame observation reveals that the formation of the chaotic area involves two distinct processes. In the first process, a few GB atoms deviate from expected positions, as highlighted by the red-coded atoms in Fig. 4b. We call these deviations the disruptive jumps, and the atoms involved are referred to as disruptive atoms. In the second process, more disruptive jumps occur until the GB completely stagnates. To analyze the energy path of this transition, we utilized the states before and after stagnation (states B and D in Fig. 4d) as the initial and final configurations, respectively, and NEB simulations. Figures 4b-e demonstrate that the NEB method could reproduce the complete disruptive jumps leading up to GB stagnation as captured by the MD simulations. Consequently, we performed multiple NEB simulations and constructed the energy landscape for GB migration, as shown in Fig. 4f.

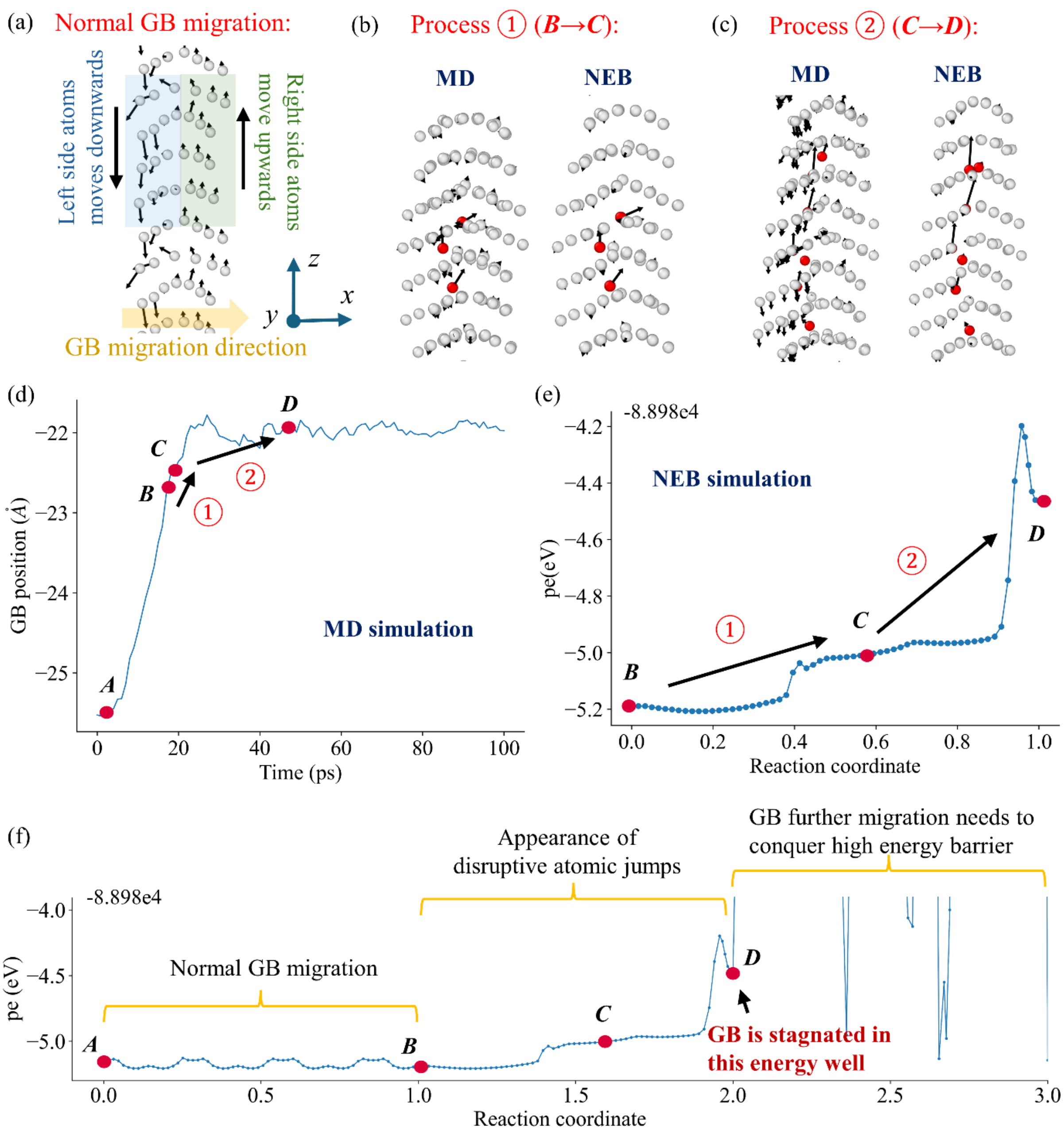

Figure 4 Detailed visualization of GB dynamics and simulations. (a) Orderly movement of GB atoms during the normal migration period. (b-c) Disruptive jumps in the GB area leading to stagnation, with a comparison of results from molecular dynamics (MD) and NEB simulations. (d) Plot of the GB position vs. time curve. (e) Minimum energy path obtained from an NEB simulation. (f) Energy landscape for GB migration.

Figure 4f illustrates that during normal GB migration, the system overcomes a series of small periodic energy barriers, allowing the GB to move rapidly. At state B, disruptive jumps by a few GB atoms occur, slowing the surrounding area and eventually causing the entire GB to stagnate.

We applied a small driving force at state B to manually move the GB beyond the position at state D without triggering disruptive jumps, using this as the final state for the NEB simulation with state D as the initial state. Figure 4f shows that further migration from state D requires overcoming extremely high energy barriers. We conducted multiple NEB simulations for this process, with the GB surpassing its position at state D by varying distances. There were only two scenarios for the GB at state D to move forward: the GB can revert to state B—a state without disruptive atoms—moving forward with lower energy barriers, or it can proceed directly, facing extremely high energy barriers. This demonstrates that GB stagnation occurs when the configuration becomes trapped in an energy well, as depicted in Fig. 4f.

But why can a high driving force induce this type of GB stagnation? There are two possible explanations. First, a high driving force can increase the velocity of GB migration, as shown in Fig. 2. This increased momentum raises the likelihood of disruptive jumps occurring among GB atoms. Second, we examined the energy barrier prior to state D by conducting NEB simulations under various shear stresses. Figure 5 demonstrates that the driving force can lower the energy barrier of the disruptive jumps of GB atoms, thereby facilitating the GB stagnation.

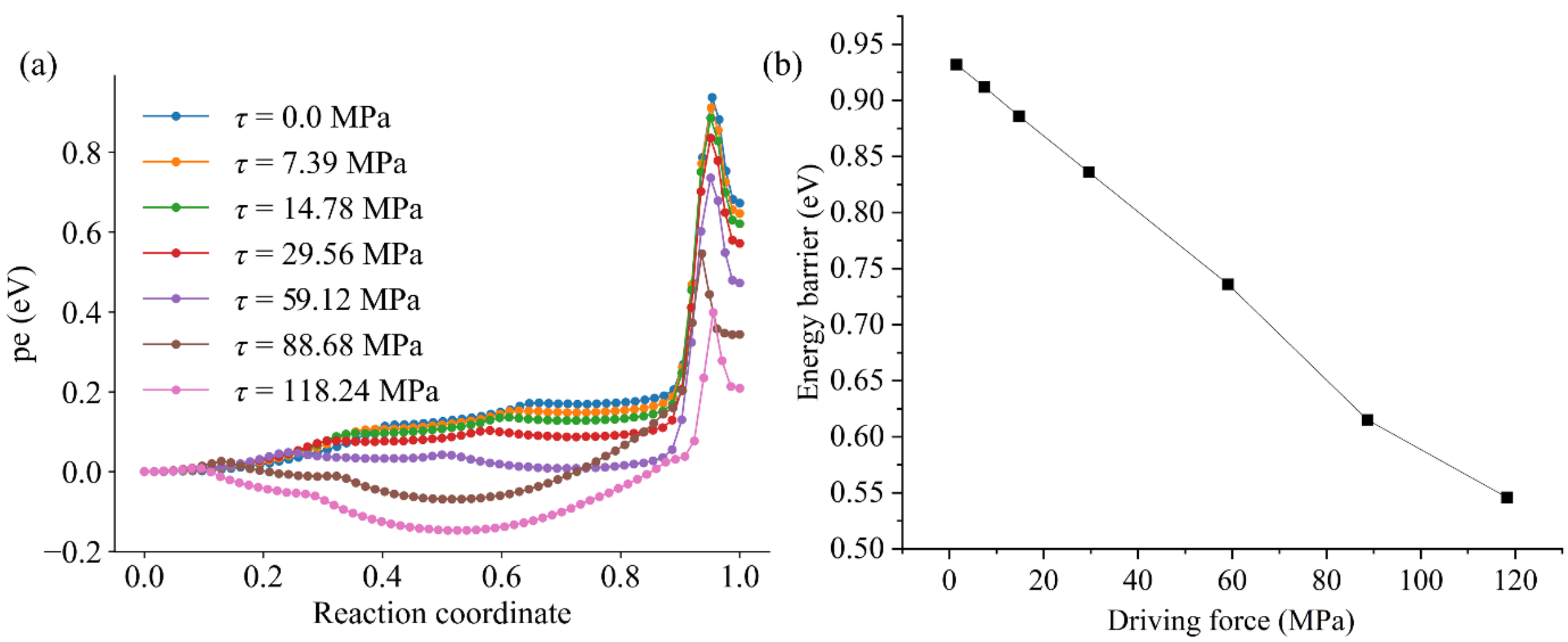

Figure 5 The effect of driving force on the energy barrier leading to the disruptive jumps in GB area

*3.2 Characterization of disruptive atoms*

The previous section illustrated that the disruptive jumps of only a few GB atoms can lead to the stagnation of the entire GB. This finding prompts an important question: how can we effectively detect these subtle disruptive jumps or disruptive atoms, especially within the GB area where long-range ordering is absent?

We calculated various properties of GB atoms, including atomic energy, volume, stresses, Voronoi tessellation [52,53], local entropy [54,55], and local mass density, techniques previously applied to characterize the local atomic environment in a complex material. However, none of these methods could accurately characterize the disruptive atoms in the GB area; see Supplemental Section S1 for detailed results.

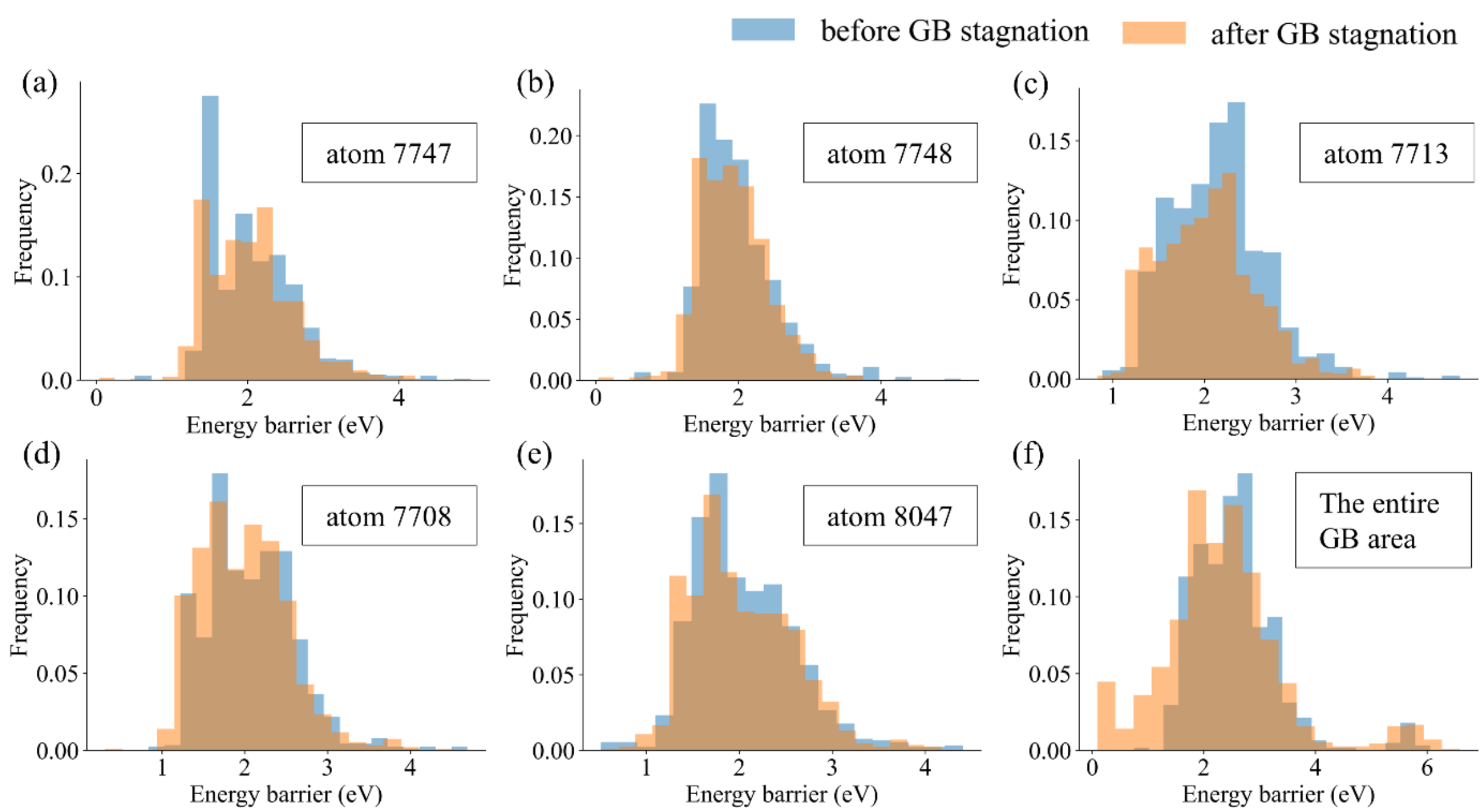

Figure 6 Energy barrier spectra before and after GB stagnation (a-e) surrounding individual GB atoms and (f) for the entire GB area. Here, the atoms 7747, 7713, 8047, 7708, and 7748 are the IDs of disruptive atoms identified by displacement vector analysis.

Additionally, we analyzed the energy barrier spectra of GB atoms before and after stagnation using the ART nouveau. According to Rodney and Schuh's study [56], the "jam transition" in the flowing metallic glass will be accompanied by a shift of energy barrier spectra to a higher energy regime, which indicate a reduction in the mobility of atoms in the metallic glass. However, in our case, no such shift in energy barrier spectra is observed, conversely, more saddle points with small energy barriers appear after the occurrence of GB stagnation (as shown in Fig. 6f). This result shows a significant difference between the GB and glass materials that, the movement of GB is a holistic action of entire structure instead of localized atomic jumps (at least before the temperature close to the melting point). That also explains why the disruptive jumps of a few GB atoms could lead to the stagnation of entire GB.

Further, we computed the displacements of all GB atoms by comparing their current positions to those from 5ps earlier and normalized the magnitudes of all displacement vectors (a displacement magnitude threshold of 0.2Å is set to minimize interference from thermal vibrations). As illustrated in Fig. 7, the vectors representing disruptive jumps differ significantly in direction from the rest. We calculated the angles between each displacement vector and all others; a vector is identified as representing a disruptive jump if the smallest angle to any other vector exceeds 0.2 radians (about 11.4592 degrees). This method effectively captures most disruptive jumps (or atoms), as shown in Fig. 7, except for those whose directions resemble those of normal atoms. To further explore more complex systems, such as those at high temperatures or involving large models, we developed a Standard Set that encompasses all normalized displacement vectors of GB atoms from the normal GB migration period (recorded from stage A to B in Fig. 4).

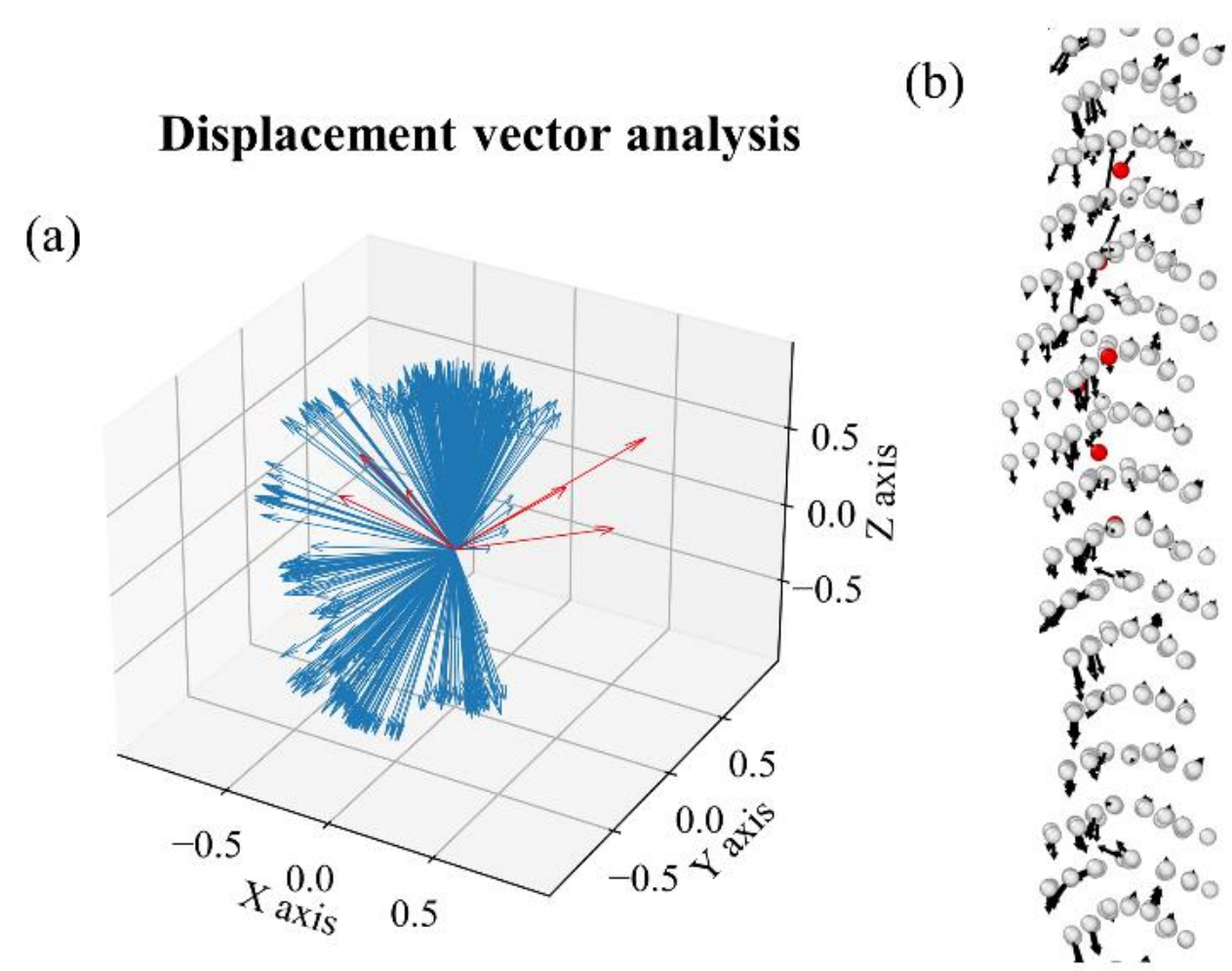

Figure 7 Distinguishing the disruptive jumps by displacement vector analysis

*3.3 Temperature induced GB stagnation*

Previous studies have demonstrated that temperatures can affect GB migration similarly to driving forces. For instance, both temperature and shear stress have been shown to induce GB structural phase transitions [57], and driving forces could lower the transition temperature of thermal behavior [58]. Therefore, we extended our investigation to examine the impact of temperature on GB stagnation to determine if similar effects occur.

Figure 8a illustrates the temperature-mobility curve of the Σ15 (2 1 1) Ni GB determined by GB random walk simulations, which exhibits anti-thermal (or non-Arrhenius) behavior, characterized by a decrease in GB mobility as temperature increases. The anti-thermal behaviour has been widely documented in both simulations [31,51,58–61] and experiments [62–65], a phenomenon have been believed inherent in the GB kinetic equation [31,58,59] or due to GB structural phase transition [25,28]. The GB displacement vs. time curves reveal that at 200K, the Σ15 (2 1 1) GB moves relatively smoothly, however, as the temperature increases to 400K, clear instances of GB stagnation are observed (Fig. 8b). Comparison of the GB structures at 200K and 400K reveals no detectable structural phase transition in the GB (Fig. 8c). According to the unified GB kinetics model proposed by Han et al. [66], GB migration can be mediated by different disconnection modes. Dichromatic analysis shows that for the Σ15 (2 1 1) GB, different disconnection modes correspond to different shear coupling factors (see Supplemental Materials Section S2 for details). The disconnection mode with the lowest activation energy, which is most likely to be activated at lower temperatures, corresponds to a shear coupling factor ($\beta$) of 0.89. Figure 8d shows consistent shear coupling factors of approximately 0.89 for GB random walks at both 200K and 400K, indicating no change in disconnection mode across these temperatures. The shear coupling behavior of Σ15 (2 1 1) Ni GB at temperatures above 500K is shown in Supplemental Fig. S7. The activation of multiple disconnection modes is likely the reason for the observed increase in mobility with rising temperatures beyond 500K.

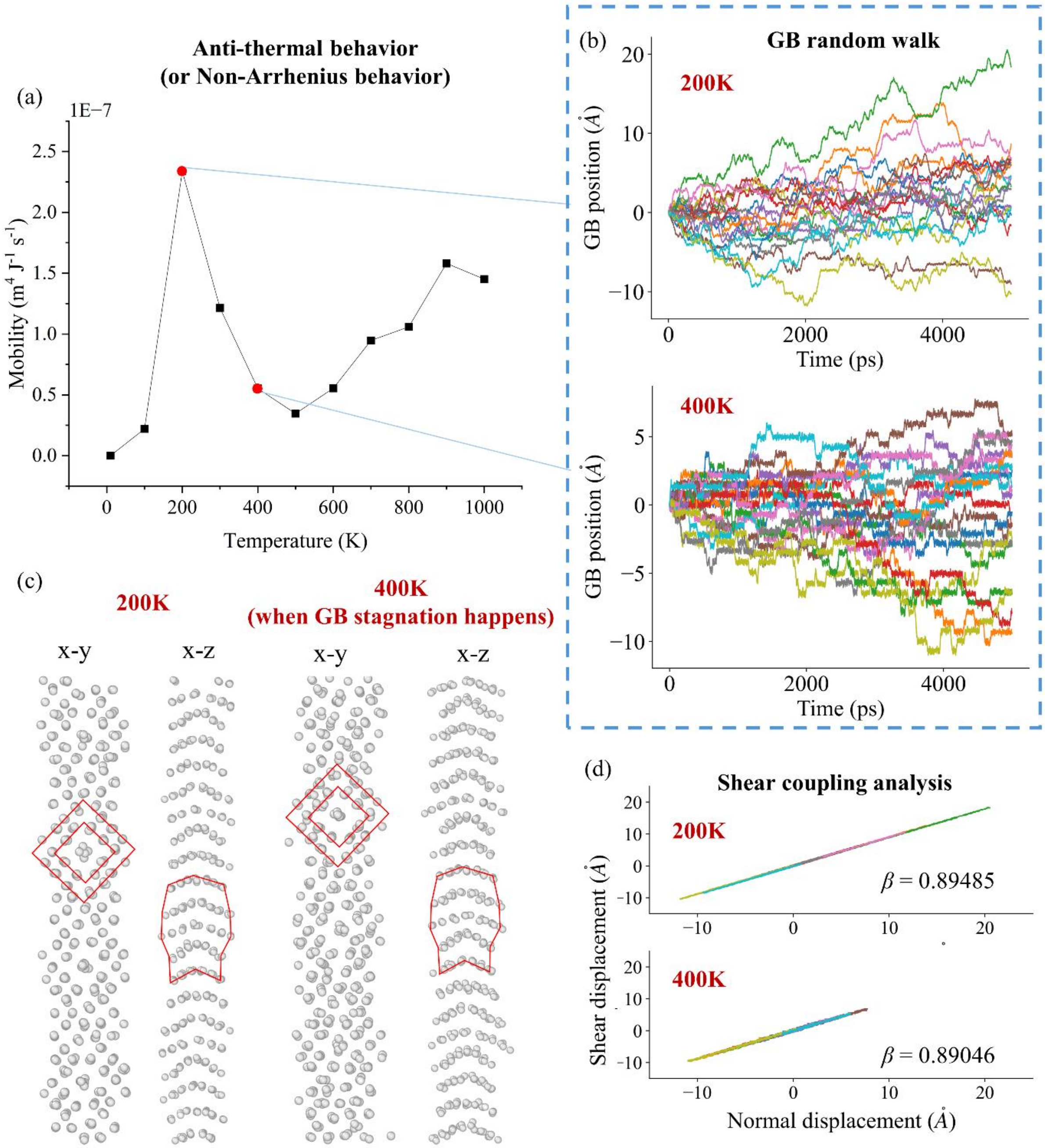

Figure 8 Results from the GB random walk simulations. (a) Mobility vs. temperature curve for the Σ15 (2 1 1) GB. (b) displacement vs. time curves, (b) structural phase analysis, and (c) shear coupling analysis of the GB at 200K and 400K. Different colors in (b) and (d) denote various independent simulation trials.

We compared the displacements of GB atoms during the random walk simulation at 400K with the Standard Set established in the previous section (here, the Standard Set includes displacement vectors for GB migration in both directions). The displacement vector analysis is applied to determine the ratio of disruptive atoms in the GB area. Figure 9 indicates that all GB stagnation stages are associated with a high ratio of disruptive jumps. The observed high ratio of disruptive jumps is similar to the roughening transition in that both are temperature-activated and disturb the arrangement pattern of GB atoms. However, unlike the increased mobility following the roughening transition noted in previous research [22,67,68], the thermally activated disruptive jumps observed here impede GB migration. In Bair and Homer's study[51] on the anti-thermal behavior of GBs, they indicate that at high temperatures, the excess volume is large, allowing some GB atoms to jump to metastable positions, which requires high energy to move their surrounding GB. Our observations confirm this phenomenon and demonstrate that these disruptive jumps can be activated not only by high temperatures but also by high driving forces.

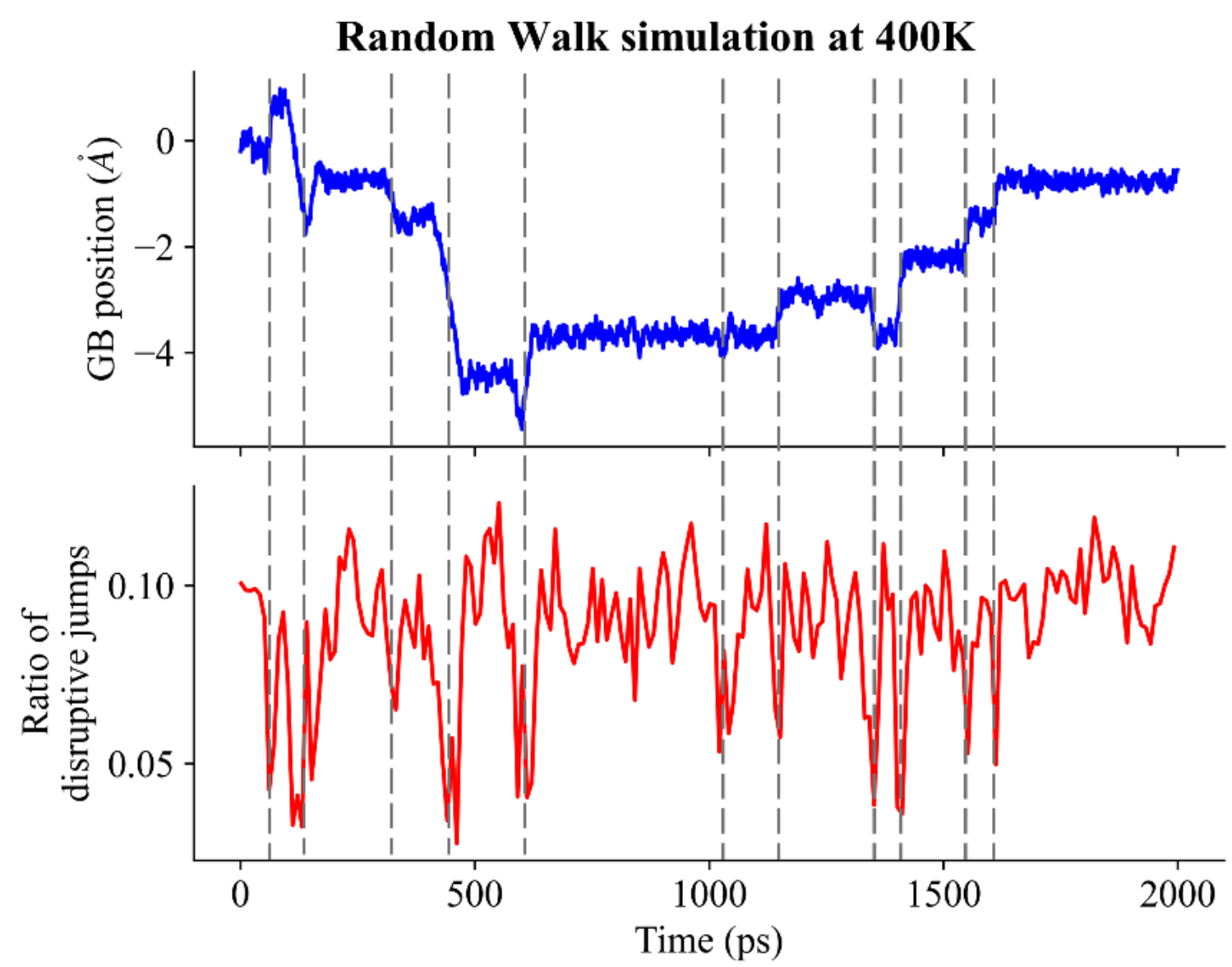

Figure 9 Variation of the ratio of the disruptive atoms in GB area during the random walk simulation at 400K

The phenomenon of significant GB stagnation at high temperatures, accompanied by lower mobility, contrasted with relatively smooth movement and higher mobility at low temperatures, is not singular. We observed similar behaviors in other GBs, such as Σ33 (7 4 1) and Σ21 (8 4 2) GBs (p43 and p76 in the Olmsted database [32]); see Supplemental Fig. S6 for details. This thermally activated GB stagnation could explain the apparent anti-thermal behavior observed in those GBs (a phenomenon we will describe as artificial due to the small model sizes in the next section). However, this mechanism does not account for all instances of anti-thermal behavior. For example, in our examination of Σ3 (11 8 5)/(11 8 $\bar{5}$), Σ7 (8 5 1)/(7 5 4), and Σ9 (7 5 4)/(7 5 $\bar{4}$) GBs (p366, p207, and p87 in the Olmsted database [32]), which have been extensively studied before [51,58,61,69], we found no evidence of thermally activated GB stagnation; see Supplemental Fig. S7 for details.

### *3.4 Effects of model size on GB stagnation and thermal behavior*

The results and discussion thus far have relied on the relatively small models (roughly 10 nm$^2$) from the Olmsted database, where the GB typically moves as a cohesive plane. In contrast, in larger models, GB migration is mediated by the nucleation and migration of disconnections [70–72], a process not present in smaller models. To better capture this complexity, we expanded the model tenfold in the z-direction, enabling a more comprehensive investigation of GB dynamics.

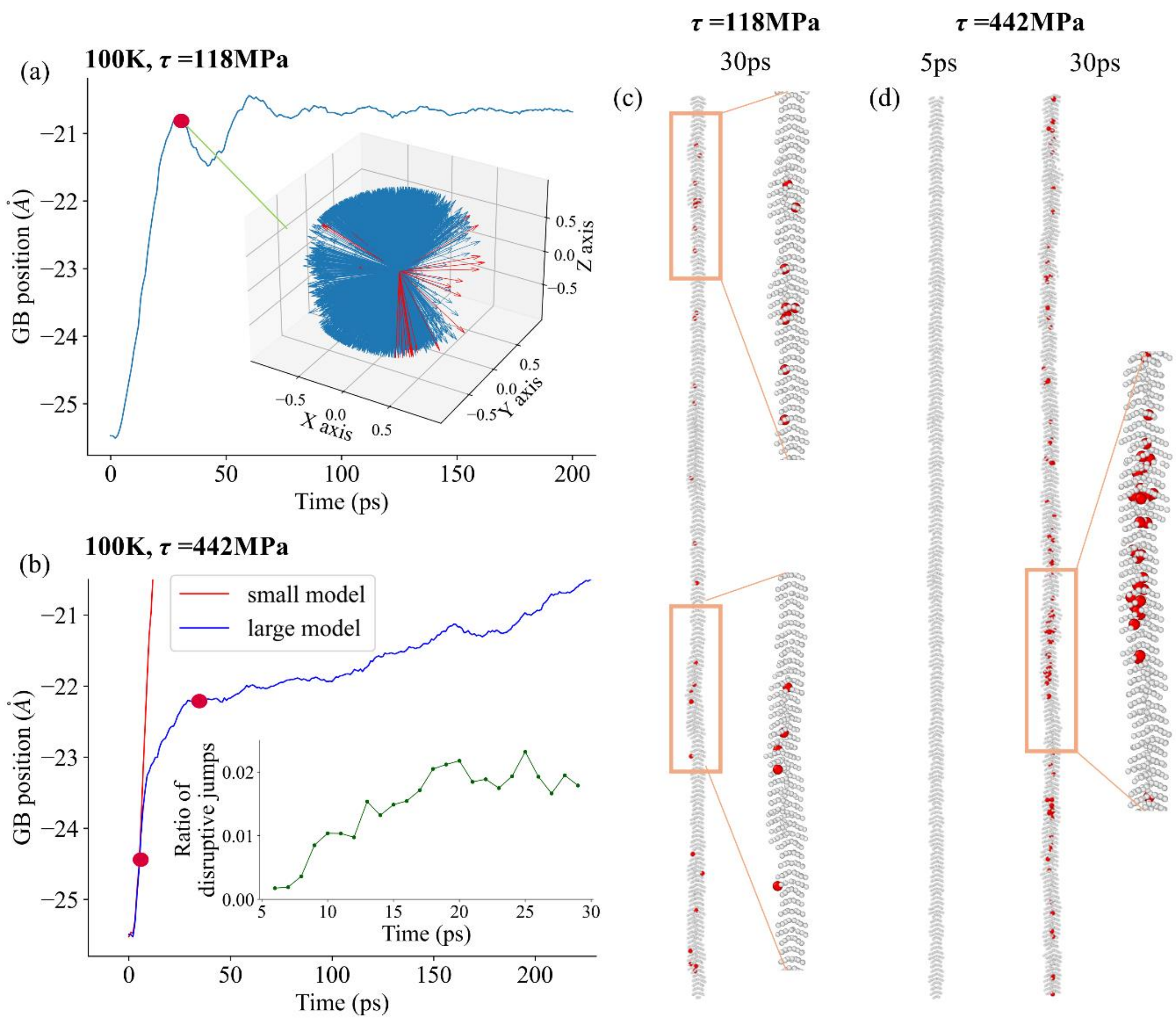

Figure 10 (a) GB stagnation at $\tau$ = 118MPa and (b) GB slowdown at $\tau$ = 442MPa in large model. Disruptive atoms are magnified twofold for clearer observation.

Figure 10a illustrates that GB stagnation can occur in large models under the same conditions as in small models. The displacement vector analysis reveals that disruptive atoms tend to accumulate in the delayed (or curved) areas. We further increased the driving force from 118 MPa to 442 MPa. After a brief period of rapid migration, the GB gradually slows down, and the ratio of disruptive atoms steadily increases to a high value. These disruptive atoms move with the GB and exert a drag effect on GB.

It is noteworthy, as shown in Fig. 10b, that during the rapid migration period, the velocity of the large model matches that of the small model. However, since the large model contains more GB atoms, it experiences a higher probability of disruptive jumps, causing its curve to gradually diverge from that of the small model as the GB progresses. Since both driving force and temperature can activate disruptive jumps, the high probability of disruptive jumps in large models may also influence the thermal behaviors of GBs. Consequently, we conducted random walk simulations of the large models for the six GBs that we have studied so far and compared their mobility-temperature curves with those from the small model. The results, illustrated in Fig.11, show a clear transition in their thermal behaviors for GBs who exhibits clear thermally activated stagnation: most transition from anti-thermal to thermally activated behavior, except for a minor peak in the Σ21 (8 4 2) GB. This transition is akin to the recently reported solute-induced shift in GB migration behavior from Non-Arrhenius to Arrhenius [3]. Therefore, the observed anti-thermal behavior in these GBs could be an artificial phenomenon due to the small model size. However, for GBs that do not exhibit obvious thermally activated stagnation, the model size appears to have no significant effect, as shown in Figs. 10(d-f) in three representative GBs.

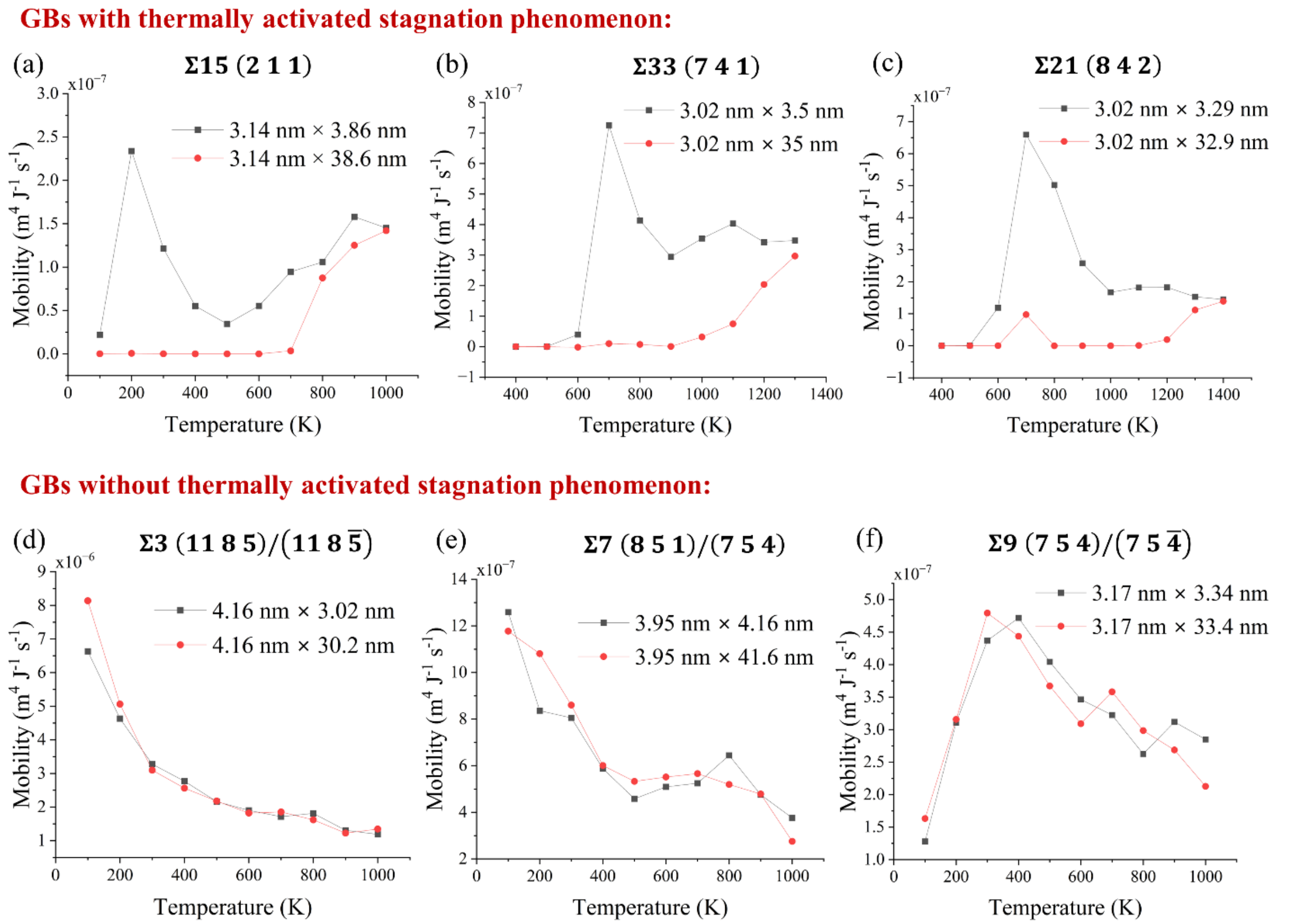

Figure 11: Comparison of thermal behavior between small and large models of different GBs. The legends indicate the dimensions of the model on the GB plane ($y$ size × $z$ size).

## 4. Discussion

The current study reveals that disruptive jumps, which interrupt the ordered GB movement pattern, can significantly reduce GB mobility, leading to GB stagnation. It is important to note that these results are based on the bicrystal model with free surfaces, which differs from real conditions in polycrystals where GB migration is constrained by triple junctions [11–14] and neighboring grains due to the shear coupling effect [38,73]. Therefore, the stagnation of a single GB cannot directly account for the stagnation of the entire GB network in polycrystals; however, they are not entirely unrelated. Holm and Foiles's study [22] demonstrates that a small ratio of low mobility GBs can lead to the cessation of the whole GB network in polycrystals. Thus, our study provides a possible mechanism for explaining the cessation of grain growth in polycrystals from a mobility perspective.

Another limitation of the current study compared to real materials is the large driving force applied in MD simulations, leading to unrealistically high GB migration velocities. Figure 2 shows that GB stagnation becomes evident when the driving force reaches 88.7 MPa and above, a value higher than the typical experimental driving force, which ranges from $10^2$ to $10^6$ Pa [74]. Detor and Schuh's study [75] demonstrates that when grain size is reduced to several nanometers, the curvature-induced driving force alone can reach $10^9$ Pa. Combined with other factors such as mechanical shear, the driving force can reach even higher magnitudes. Therefore, disruptive jumps could be more prevalent in nanocrystalline materials than in coarse-grained materials. Despite this, the high velocity (~100 m/s) observed in the current study is unlikely in real materials. Experimental surveys on a limited number of GBs across the five-parameter GB space show velocities in the range of $10^{-8}$ to $10^{-3}$ m/s [74]. However, previous computational simulations of extensive GB databases reveal many GBs, particularly Σ3, Σ7, and Σ9 GBs [51,60], with extremely high mobility even at low temperatures. For example, the Σ3 (11 8 5)/(11 8 $\bar{5}$) GB, which has been extensively studied [61,69], exhibits a high mobility of 8.14 m/(s MPa) at a low temperature of 100K, indicating it can easily achieve high velocities even under experimental conditions. Verma et al.'s study [69] and Fig. 11d demonstrate that this high mobility is not an artifact of the small model size in MD simulations. Whether these high-mobility GBs exist in the real world remains debated.

A possible explanation for these discrepancies between MD simulation and real world is the difference in event frequency between small and large scales[76,77]. Deng et al.'s experiments[77] show that a fresh cell can response immediately and exhibit fast move after being subjected to a driving force for a short time (0-0.01 s). However, over a longer timescale, the cell exhibits slow creep (0.01-10 s). They attribute this timescale-dependent behavior to thermal fluctuations in semiflexible polymers[78,79]. This principle could be extended from fresh cell to GBs. Additionally, constraints in real materials such as triple junction drag and shear coupling effect during GB migration could also prevent fast GB migration. Figure 2 shows that when the driving force is large enough, disruptive jumps can occur in a very short time. Further investigation is needed to determine whether these disruptive jumps can occur before the constraints on GB migration in polycrystals become significant.

The following is a possible scenario in polycrystals during grain growth. Initially, the GB exhibits significant curvature due to the small grain size, and substantial residual stresses from mechanical deformation are present around the GB. These factors create a large driving force, facilitating the occurrence of disruptive jumps in the GB area. Initially, this large driving force can overcome the drag effect of disruptive jumps, allowing the GB to migrate slowly forward (Fig. 10b). As GB migration progresses, the associated residual stresses diminish, and the curvature of the GB decreases due to grain growth. Consequently, when the driving force decreases to a certain threshold, the drag effect of the disruptive jumps becomes predominant, leading to the stagnation of GB movement. Nevertheless, as shown in Supplemental Fig. S1f, the rearrangement of the GB structure may occur over an extended period. While the GB can move again due to this rearrangement, it soon stagnates due to the large driving force. In real-world material relaxation timescales, typically spanning hours or days, GB rearrangement may occasionally happen. When the driving force reduces to a value insufficient to trigger disruptive jumps, the GB can exhibit continuous motion. This is a hypothetical process, and further investigations are needed to validate it.

## 5. Conclusion

Through atomistic simulations, we have unveiled a mechanism for GB stagnation in pure materials and non-Arrhenius migration behaviors. The main findings are as follows:

1. The disruptive jumps within the GB area can break the ordered movement pattern of GB migration, significantly reducing GB mobility. Remarkably, even disruptive jumps of just a few atoms can lead to the stagnation of the entire GB.

2. This type of disruptive jumps can be triggered by either high driving forces or elevated temperatures, leading to GB stagnation.

3. Increasing the model size enhances the probability of disruptive jumps. We also observed a transition in the thermal behavior of GB migration from anti-thermal to thermally activated as the model size increased. However, model size has a negligible effect on GBs that do not exhibit evident thermally activated stagnation.

4. The disruptive jumps and associated atoms are indistinguishable from other GB atoms in terms of atomic energy, volume, local density, local entropy, or Voronoi tessellation, and no “jam transition” is observed in the energy barrier spectra before and after the GB stagnation. This makes detection of disruptive jumps in GB area challenging. We propose a displacement vector analysis method that effectively capture these subtle disruptive jumps within the GB area.

The observations expand our understanding of self-caused GB stagnation mechanisms and provides a possible explanation for why GBs cannot continuously grow in materials during heat treatment, even in the absence of impurities.

**Acknowledge**

The authors thank Dr. Normand Mousseau for sharing the ART nouveau codes, and Dr. David L Olmsted for sharing the 388 Ni GB structure database. This research was supported by NSERC Discovery Grant (RGPIN-2019-05834), Canada, and the use of computing resources provided by Research Alliance of Canada. X.S. also acknowledges financial support from the University of Manitoba Graduate Fellowship (UMGF). During the preparation of this manuscript the authors used ChatGPT to improve its readability. After using this tool, the authors reviewed and edited the manuscript as needed and take full responsibility for the content of the publication.

**Supplementary materials**

**for**

# Disruptive Atomic Jumps Induce Grain Boundary Stagnation

Xinyuan Song and Chuang Deng*

Department of Mechanical Engineering, University of Manitoba, Winnipeg, MB R3T 2N2, Canada

* Corresponding author: Chuang.Deng@umanitoba.ca

## S1. Long-term driving force driven grain boundary (GB) migration simulations at 100K under different magnitude of shear stresses

Figure 2 in the main article illustrates the phenomenon of high driving force-induced grain boundary (GB) stagnation. For clarity, we present a short-term representative result for each driving force. As supplemental information, Figure S1 displays results from long-term simulations, each lasting 5 ns. At each driving force, the simulation is repeated 20 times with different initial velocity distributions.

When the shear stress is low (7.39 MPa to 14.8 MPa), the GB movement pattern resembles the "stick-slip" migration pattern reported in Bair and Homer's work [1], where GBs intermittently stop and then continue moving. At a shear stress of 59.1 MPa, while most GBs continue moving and reach the model boundary, one stagnates in the middle and remains stationary until the end of the simulation. At 88.7 MPa, most GBs experience long-term stagnation, with only a few reaching the model boundary. At 118.2 MPa, all GBs stagnate after migrating a short distance. This phenomenon has been noted in other literature [2].

The observed phenomenon is a form of long-term stagnation. GB stagnation first occurs at a shear stress of 59.1 MPa, and its frequency increases with higher shear stress. Results from 20 repeated simulations clearly demonstrate the increasing frequency and effect of the driving force. At 118.2 MPa, GB stagnation occurs in all simulations within a very short time, indicating that this phenomenon is not uncommon for Σ15 (2 1 1) Ni GB at high driving forces.

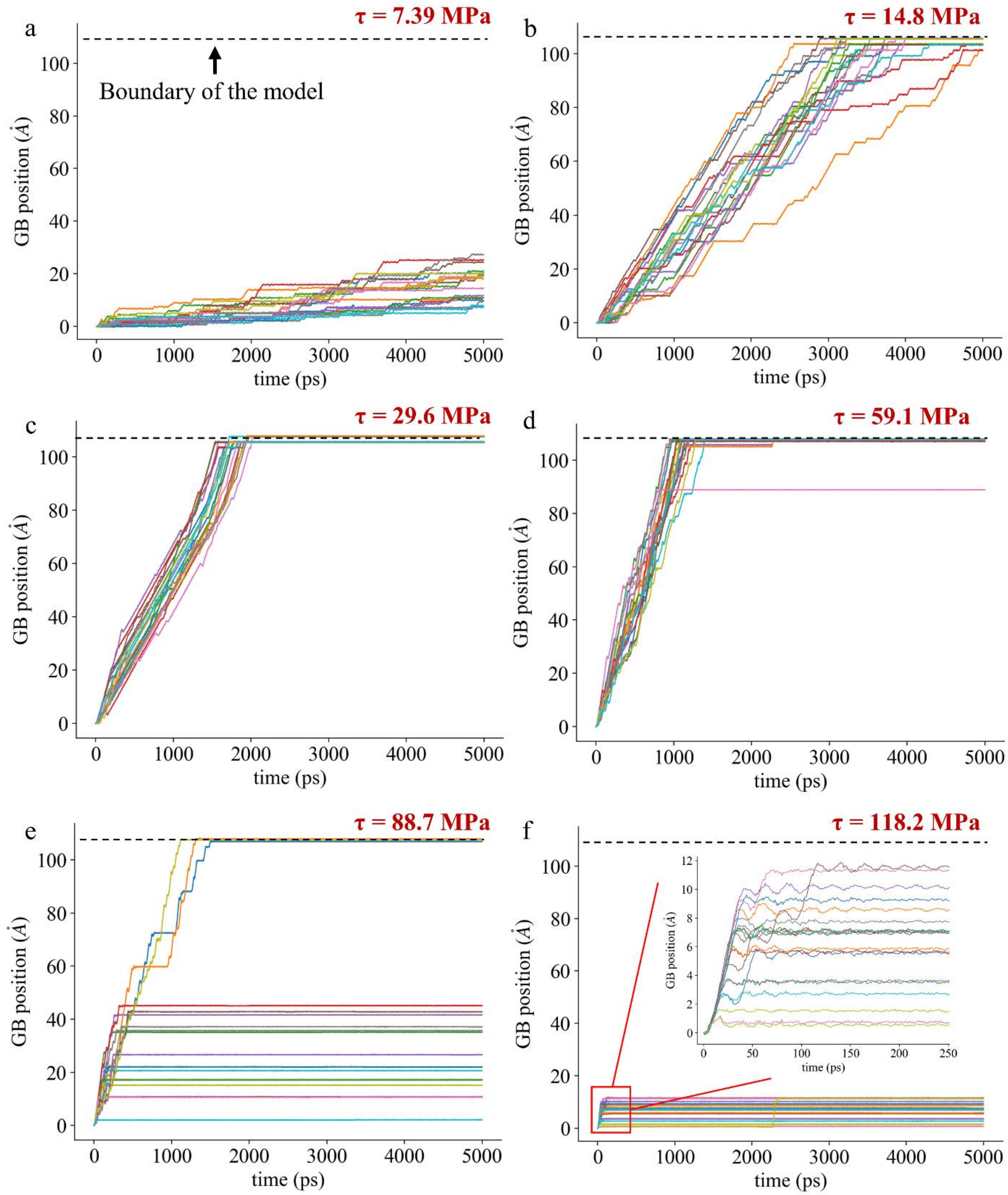

Figure S1 Plots of GB position vs. time of Σ15 (2 1 1) GB under different shear stresses at 100K. Different colored curves represent different simulations with varying initial velocity distributions.

**S2. Characterizing the disruptive atoms by different descriptors**

In the main article, we demonstrated that the disruptive jumps of a few atoms, referred to as disruptive atoms, can lead to the stagnation of the entire grain boundary (GB). In this section, we evaluated various descriptors to characterize these disruptive atoms. Figure S2 illustrates the distribution of atomic energy in GB following stagnation. Displacement vector analysis identifies the disruptive area and atoms. However, when compared to normal areas and GB atoms, the disruptive area and atoms exhibit no significant differences.

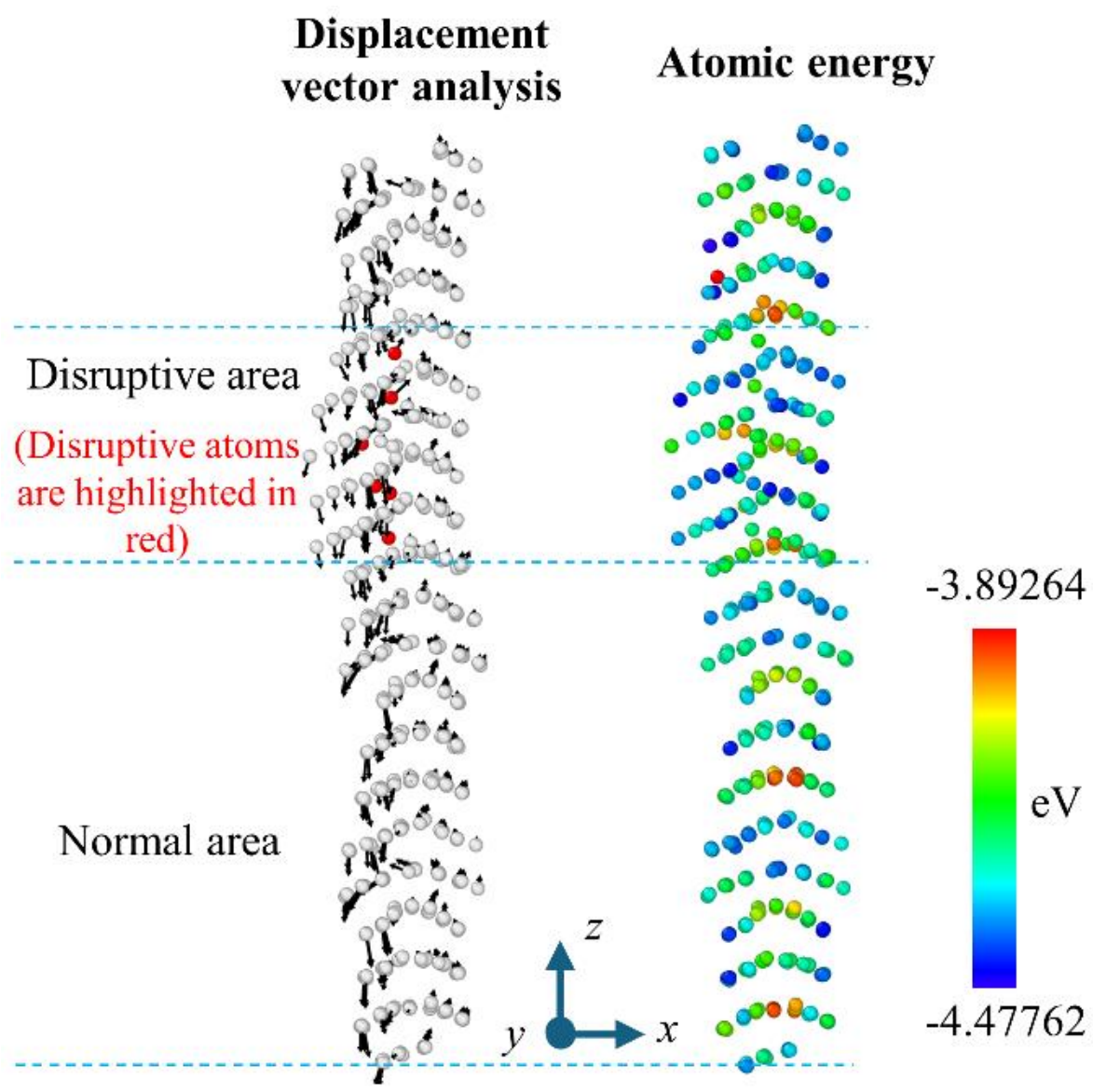

Figure S2 Distribution of atomic energy in GB area. Displacement vector analysis identifies the disruptive area and atoms.

We subsequently analyzed the Voronoi tessellation [3] in the GB area following GB stagnation. Figure S3 depicts the distribution of atomic volume (Voronoi volume) and the neighbor count of Voronoi cells; however, neither parameter effectively reflects the disruptive area or disruptive atoms. Table S1 enumerates the ten most frequent Voronoi indices in the GB area before and after GB stagnation. According to deviations reported in the literature [4], the frequency of crystal-like clusters (0 4 4 x) shifted slightly from 6.2% to 6.4%, and icosahedral-like clusters (0 2 8 x) decreased from 2% to 1.7%, while mixed clusters (0 3 6 x) increased from 5.1% to 7.5%. These minor changes are insufficient to characterize GB stagnation.

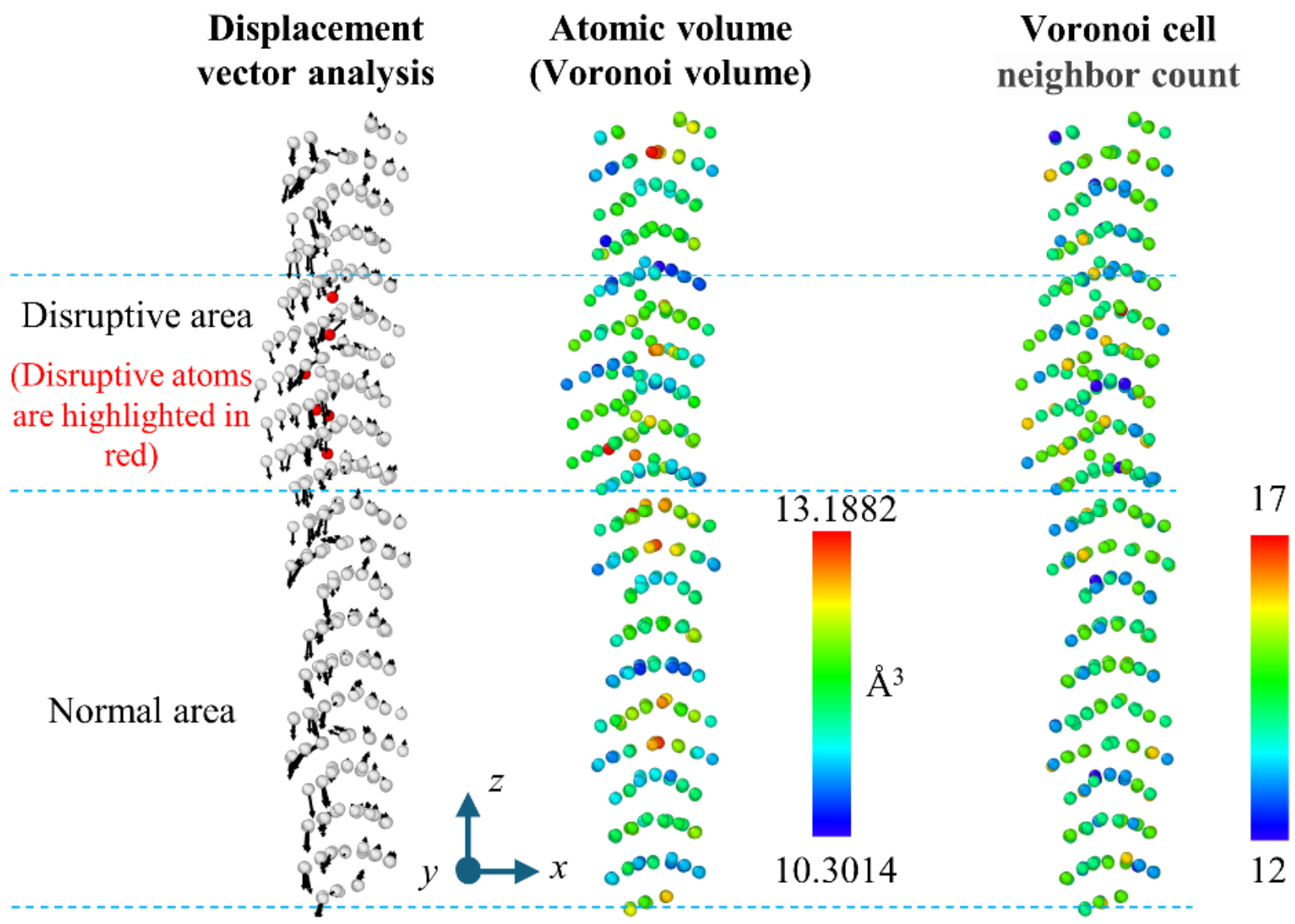

Figure S3 Distribution of atomic volume and neighbor count within Voronoi cells in the grain boundary (GB) area. The neighbor count represents the number of Voronoi cells connected to a given cell, which can also be interpreted as the number of faces on each Voronoi cell. Displacement vector analysis helps identify the disruptive area and atoms.

Table S1 Lists the ten most frequent Voronoi indices in the GB area before and after GB stagnation

| Before GB stagnation | | After GB stagnation | |
|---|---|---|---|
| Voronoi indices | frequency | Voronoi indices | frequency |
| 0 3 6 3 | 3.7 % | 0 3 6 3 | 5.8 % |
| 0 4 4 4 | 3.0 % | 0 4 4 4 | 3.5 % |
| 1 3 5 3 | 2.2 % | 0 4 4 3 | 2.9 % |
| 0 2 8 2 | 2.0 % | 0 4 6 2 | 2.7 % |
| 0 4 4 3 | 1.6 % | 1 3 5 2 | 2.5 % |
| 0 4 4 2 | 1.6 % | 0 3 6 4 | 1.7 % |
| 0 5 4 3 | 1.4 % | 0 2 8 2 | 1.7 % |
| 1 1 7 3 | 1.4 % | 0 5 4 3 | 1.5 % |
| 0 5 3 3 | 1.4 % | 0 4 5 3 | 1.5 % |
| 0 3 6 2 | 1.4 % | 0 4 6 1 | 1.3 % |

Further investigation, including assessments of local mass density and local entropy (Fig. S4) [5,6], as well as atomic stresses in various directions (Fig. S5), has been conducted. However, none of these parameters effectively characterize the disruptive area or disruptive atoms within the GB area.

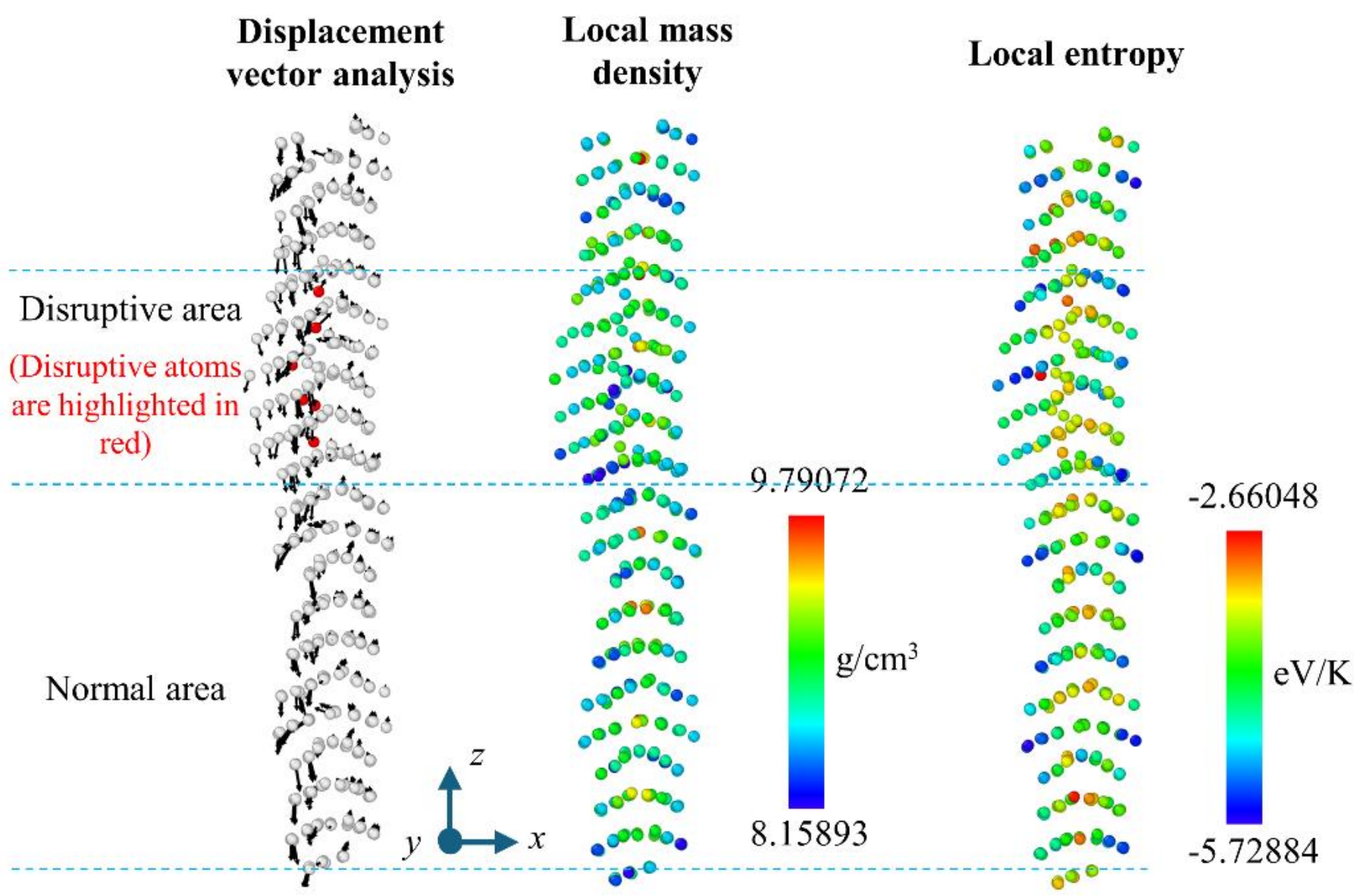

Figure S4 Distribution of local mass density and local entropy in GB area. Displacement vector analysis identifies the disruptive area and atoms.

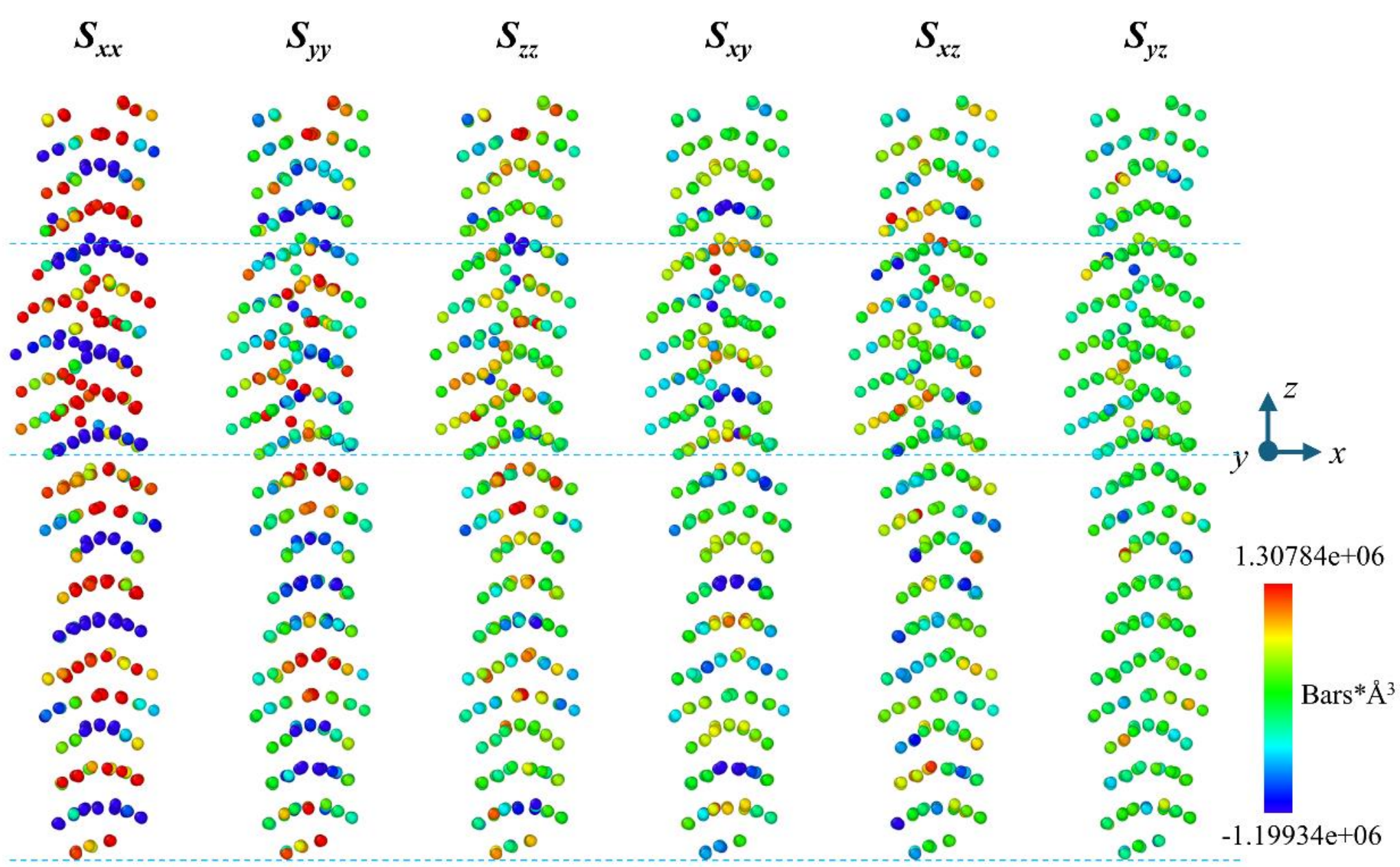

Figure S5 Distribution of atomic stresses in different directions in GB area.

**S3. Dichromatic analysis of Σ15 (2 1 1) Ni GB and the change in shear coupling behavior after 500K**

According to the unified GB kinetics model proposed by Han et al. [7], GB migration is mediated by different disconnection modes. Disconnections are line defects characterized by height ($h$) and Burgers vector ($\boldsymbol{b}$). The activation energy of disconnections rises with increases in the magnitude of $\boldsymbol{b}$ or h. The dichromatic pattern, a visualization technique, is achieved by extending the grains on either side of the GB throughout space. This extension facilitates the analysis of different disconnection modes through displacing the atoms on one side of the GB by a vector of $\boldsymbol{b}$ in the dichromatic pattern to restore the coincidence-site lattice (CSL) pattern, and the GB will move a distance of $h$ accordingly.

Figure S6 shows that for Σ15 (2 1 1) GB, different disconnection modes exhibit different shear coupling factors ($\beta$). Therefore, changes in disconnection mode can be identified by monitoring shear coupling variations during simulation. Notably, the disconnection mode with the lowest activation energy, corresponding to the shortest $\boldsymbol{b}$ and $h$, exhibits a shear coupling factor of 0.89.

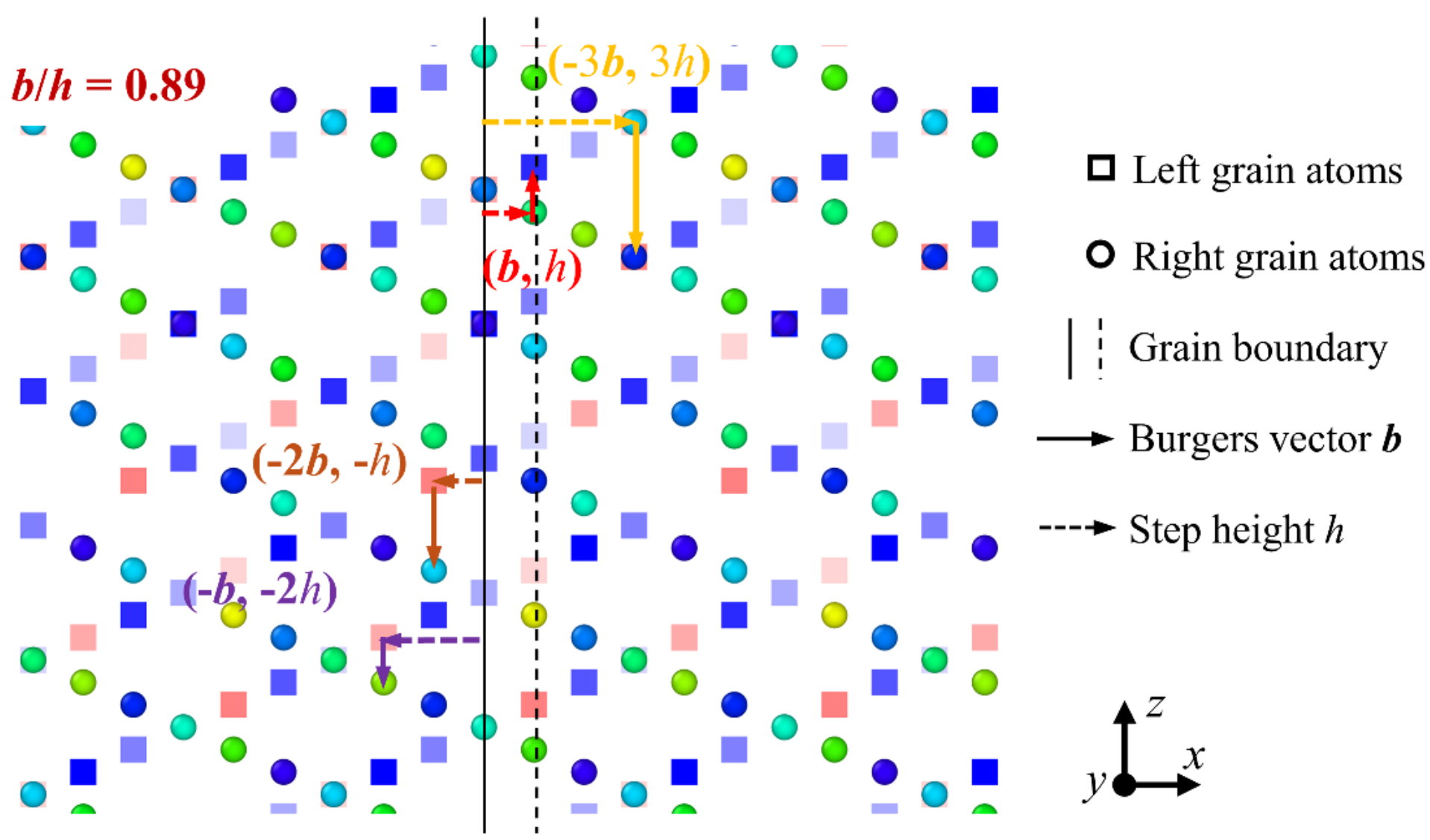

Figure S6 Dichromatic analysis of Σ15 (2 1 1) GB

Figure 7 illustrates the shear coupling behavior of Σ15 (2 1 1) GB at different temperatures. Results are obtained from Random Walk simulations. When the temperature is low (400K and below), only one disconnection mode is activated, and no phase transition is detected. Consequently, the thermally activated disruptive jumps in the GB area can cause significant GB stagnation and substantially reduce GB mobility. This observation explains the abnormal trend of GB mobility below 400K. However, as the temperature rises to 500K, Figure S7b shows that some GB normal-shear displacement curves deviate from the original path, indicating the activation of additional disconnection modes. With further temperature increases, more disconnection modes are activated, facilitating GB migration and increasing its mobility. This trend is consistent with the observed increase in GB mobility beyond 500K. Despite multi-mode mediated migration at high temperatures, GB mobility remains significantly lower than at 200K, suggesting that the drag effect of thermally activated disruptive jumps persists, albeit compensated by other mechanisms at elevated temperatures.

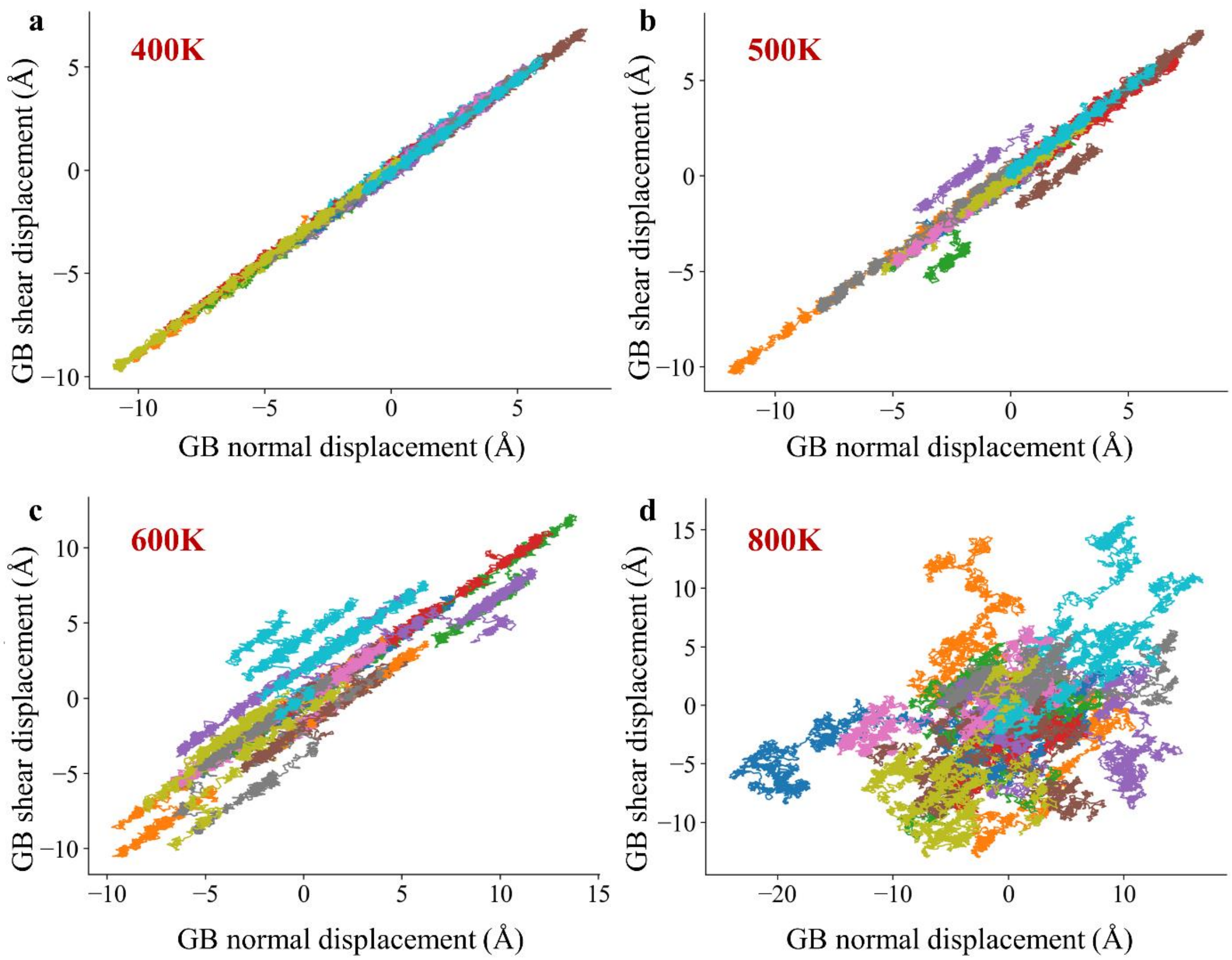

Figure S7 Shear coupling behavior of Σ15 (2 1 1) GB at different temperatures. Results are obtained from Random Walk simulations, with each colored curve representing a different independent simulation.

## S4. The generality of the thermally activated GB stagnation

In addition to the Σ15 (2 1 1) GB discussed in the main article, we have observed thermally activated GB stagnation in other GBs, such as Σ33 (7 4 1) and Σ21 (8 4 2) GBs, as illustrated in Fig. S8. This thermally activated GB stagnation can account for the anti-thermal behavior observed in these GBs. However, not all GBs exhibiting anti-thermal behavior show evident thermally activated GB stagnation, such as the Σ3 (11 8 5)/(11 8 $\bar{5}$), Σ7 (8 5 1)/(7 5 4), and Σ9 (7 5 4)/(7 5 $\bar{4}$) GBs, as shown in Fig. S9.

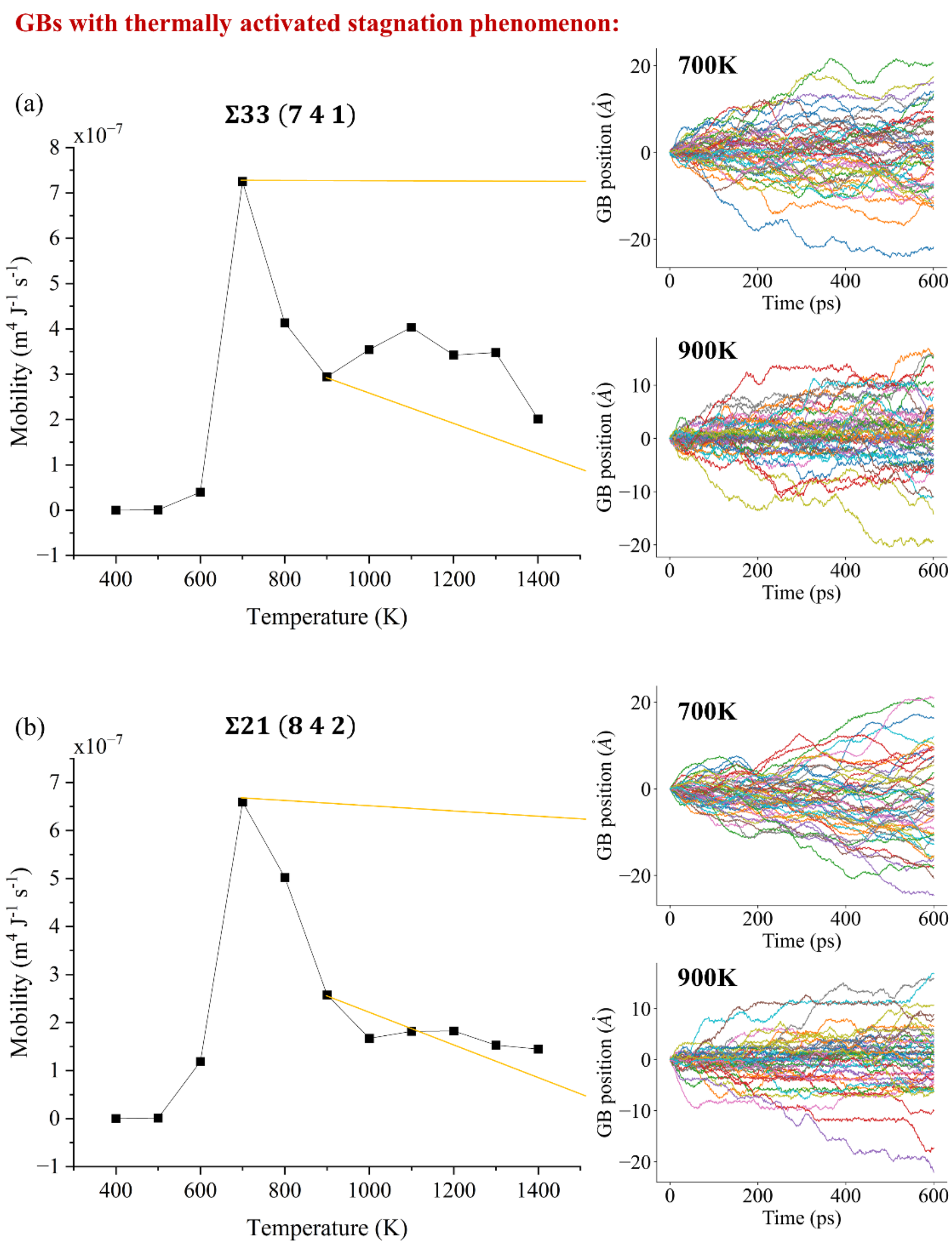

Figure S8 Mobility-Temperature curves of (a) Σ33 (7 4 1) and (b) Σ21 (8 4 2) GBs. The accompanying Displacement-Time curves clearly demonstrate thermally activated GB stagnation

**GBs without thermally activated stagnation phenomenon:**

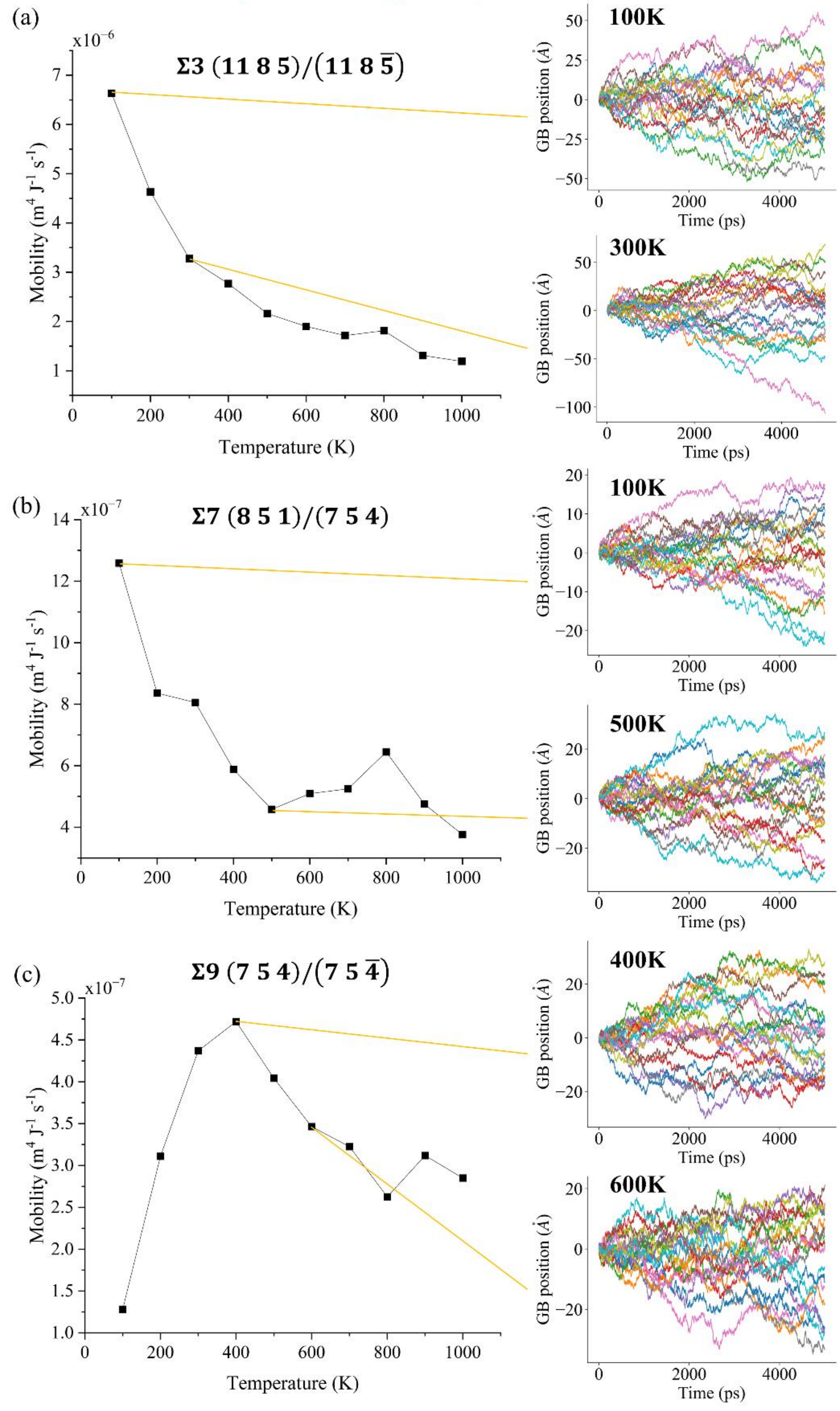

Figure S9 Mobility-Temperature curves of (a) Σ3 (11 8 5)/(11 8 $\bar{5}$), (b) Σ7 (8 5 1)/(7 5 4), and (c) Σ9 (7 5 4)/(7 5 $\bar{4}$) GBs. No evident thermally activated GB stagnation is observed in their Displacement-Time curves.